# Protonation-induced discrete superconducting phases in bulk FeSe single crystals

Yan Meng[a,b,1], Xiangzhuo Xing[a,c,1,*], Xiaolei Yi[a,1], Bin Li[d], Nan Zhou[a], Meng Li[a], Yufeng Zhang[a], Wei Wei[a], Jiajia Feng[a], Kensei Terashima[b], Yoshihiko Takano[b], Yue Sun[a,e,*], and Zhixiang Shi[a,*]

[a] *School of Physics, Southeast University, Nanjing 211189, China*

[b] *International Center for Materials Nanoarchitectonics (MANA), National Institute for Materials Science, 1-2-1 Sengen, Tsukuba, Ibaraki 305-0047, Japan*

[c] *School of Physics and Physical Engineering, Qufu Normal University, Shandong 273165, China*

[d] *Information Physics Research Center, Nanjing University of Posts and Telecommunications, Nanjing 210023, China*

[e] *Department of Physics and Mathematics, Aoyama Gakuin University, Sagamihara 252-5258, Japan*

[*] Corresponding authors：xzxing@qfnu.edu.cn, sunyue@phys.aoyama.ac.jp, and zxshi@seu.edu.cn

[1] These authors contributed equally to this work.

## Abstract

The superconducting transition temperature, $T_c$, of FeSe can be significantly enhanced several-fold by applying pressure, electron doping, intercalating spacing layer, and reducing dimensionality. Various ordered electronic phases, such as nematicity and spin density waves, have also been observed accompanying high-$T_c$ superconductivity. Investigation on the evolution of the electronic structure with $T_c$ is essential to understanding electronic behavior and high-$T_c$ superconductivity in FeSe and its derived superconductors. In this report, we have found a series of discrete superconducting phases, with a maximum $T_c$ up to 44 K, in $H^+$-intercalated FeSe single crystals using an ionic liquid gating method. Accompanied with the increase of $T_c$, suppression of the nematic phase and evolution from non-Fermi-liquid to Fermi-liquid behavior was observed. An abrupt change in the Fermi surface topology was proposed to explain the discrete superconducting phases. A band structure that favors the high-$T_c$ superconducting phase was also revealed.

# 1 Introduction

The iron chalcogenide superconductor FeSe has recently attracted tremendous interest owing to its exotic properties [1,2], including the tunability of its superconducting (SC) transition temperature, $T_c$. FeSe shows bulk superconductivity with a $T_c$ of ~ 9 K [3], which is accompanied by nematic electronic order below the structural transition at $T_s \approx 87$ K and in the absence of long-range magnetic order. Upon applying pressure, the nematic phase is suppressed, and a new magnetic order is stabilized at low pressures. $T_c$ can be significantly enhanced up to ~ 40 K when the magnetic order vanishes at high pressures [4]. These results indicate that the suppression of nematicity and the presence of magnetism may be a prerequisite for higher $T_c$. Meanwhile, superconductivity, nematicity, and magnetism compete with each other in pressurized FeSe. In addition to the application of pressure, bulk superconductivity with $T_c$ above 30–40 K has also been achieved in heavily electron-doped FeSe by intercalating alkali/alkali-earth metals, ammonia, or organic molecules into the adjacent FeSe layers [5-18]. More surprisingly, signs of superconductivity with $T_c$ exceeding 65 K have been observed in monolayer FeSe/SrTiO$_3$; strong interfacial electron–phonon coupling and charge transfer through the interface have been suggested to be crucial for this high-$T_c$ superconductivity [19,20]. Subsequently, high-$T_c$ superconductivity was also induced in K-dosed FeSe films [21,22]. In sharp contrast to bulk FeSe, whose Fermi surface (FS) consists of both hole and electron pockets [1], a common FS topology with only electron pockets in the Brillouin zone corners has been demonstrated in FeSe-derived high-$T_c$ superconductors [23-26], monolayer FeSe/SrTiO$_3$ [27,28], and K-dosed FeSe films [21,22], revealing that electron doping plays an important role in achieving high-$T_c$ superconductivity.

The electrostatic gating technique [29,30], which can systematically tune carrier concentration, is an effective way to induce high-$T_c$ superconductivity and study $T_c$ evolution in FeSe [31-36]. Recently, an unusual and discrete SC phase diagram [35] was mapped out through intercalating Li$^+$/Na$^+$ cations into FeSe flakes by a solid ionic gating device [34]. Such discrete SC phase diagrams are quite different from those that are dome-shaped and have been commonly established in unconventional superconductors. However, these discrete diagrams are in accordance with observations of certain $T_c$s in previously reported intercalated FeSe superconductors. As the carrier concentrations increase, a Lifshitz transition or an abrupt change in pairing strength is proposed to be responsible for the sudden growth of $T_c$ [33,35]. Thus, the evolution of the electronic structure is essential to understanding discrete high-$T_c$ SC phases as well as their physical properties. However, the electron doping induced by intercalation in FeSe thus far is always concomitant with the change of crystal structure, from

"11" anti-PbO-type to "122" ThCr$_2$Si$_2$-type [5-17,34,35], complicating the understanding of the electronic structure evolution and the $T_c$ enhancement mechanism. Moreover, along with the discrete SC phases, it is of great importance to determine how the electronic orders and electronic properties evolve with doping. These parameters can provide key insights for achievement of high-$T_c$ superconductivity.

In this paper, we systematically study evolution of the normal state transport property and superconductivity with electron doping in bulk H$_x$-FeSe single crystals using an ionic liquid gating method with induced H$^+$ intercalation, which has negligible effect on the crystal structure [37]. In addition to the $T_c \approx 10$ K SC phase of pristine FeSe, three subsequent SC phases with different $T_c$ values emerge successively with increasing carrier concentration. Such discrete SC phases with an opposite trend of $T_c$ are also observed in the deprotonation process, wherein the amount of H$^+$ intercalation is gradually decreased. Our results definitively prove that the discrete SC phases are intrinsic and intimately related to the carrier concentrations, and the interlayer distance is not a primary factor for achieving high-$T_c$ superconductivity. Moreover, the evolution of both the normal state transport property and superconductivity are also discussed in terms of the change in FS topology induced by electron doping.

## 2 Experimental details

Pristine FeSe single crystals were grown by the chemical vapor transport method [38,39]. H$_x$-FeSe single crystals were obtained using ionic liquid gating method induced protonation, following Refs. [37,40]. A schematic illustration of the protonation technique is shown in Fig. 1. The pristine FeSe single crystal used for protonation is cleaved down to 15 ±2 μm in thickness. H$_x$-FeSe single crystal with a single 44 K SC phase (protonated for 20 days) was selected to study the evolution of $T_c$ in the deprotonation process. Details regarding the protonation/deprotonation process are described in the Supplementary Material. H$_x$-FeSe single crystals with different protonation and deprotonation times are respectively labelled with SPD and SDPD, where D stands for the protonation or deprotonation times (days). Considering that the amount of H$^+$ intercalation (the value of $x$) cannot be clearly quantified, we hereafter instead use the protonation or deprotonation time to represent the carrier concentration. Powder X-ray diffraction measurements were performed at room temperature on a commercial Rigaku diffractometer with Cu $K\alpha$ radiation. The faces of the crystals were oriented parallel to flat sample holder to give only (00$l$) diffraction peaks. Single crystal X-ray diffraction measurements were performed on a Bruker APEX-II CCD diffractometer with Mo

$K\alpha$ radiation. Data collection was performed at 150 K. Using Olex2 [41], the structure was solved with the SHeLXT structure solution program [42] and refined with the SHeLXL refinement package using Least Squares minimization [43]. Electrical transport and magnetization measurements were performed on a physical property measurement system (PPMS-9T, Quantum Design) and a magnetic property measurement system (MPMS-5T, Quantum Design), respectively.

## 3 Results and discussion

Figs. 2(a)–(c) show the evolution of the temperature dependence of the normalized resistivity, $\rho(T)/\rho_{200\ K}$, with protonation time. In pristine FeSe single crystal, the nematic transition occurs near $T_s \approx 89$ K and is accompanied by a SC transition with onset $T_{c1} \approx 10$ K, in agreement with previous reports [1,38,39]. With increasing protonation time, $T_s$ is gradually decreased to lower temperatures and disappears at the fifth day. Meanwhile, a new SC phase with onset $T_{c2} \approx 25$ K appears and then becomes sharper and sharper with protonation [see Fig. 2(a)]. This indicates competition between nematicity and high-$T_c$ superconductivity, consistent with the result of high pressure study [4]. By contrast, it is found that the nematic state is irrelevant to the emergent high-$T_c$ superconductivity in Na$_{0.35}$(C$_3$N$_2$H$_{10}$)$_{0.426}$Fe$_2$Se$_2$ [44]. Such discrepancy may be related to the peculiar orthorhombic structure that includes the ordered 1,3-diaminopropane molecules in Na$_{0.35}$(C$_3$N$_2$H$_{10}$)$_{0.426}$Fe$_2$Se$_2$. As the protonation time is further increased, another two subsequent SC transitions with onset $T_{c3} \approx 34$ K and $T_{c4} \approx 44$ K appear successively, coexisting with the SC transition at 25 K in a certain protonation time range [see Fig. 2(b)]. Then, both the 25 K and 34 K SC phases vanish, and only the 44 K SC phase persists after 15 days [see Fig. 2(c)]. To check whether the superconductivity induced by protonation is a bulk characteristic, we also performed the magnetization measurements on three typical samples (SP8, SP12, and SP20) with different protonation time, as shown in Fig. S2 in the Supplementary Material. The values of $T_c$ determined by the diamagnetic signal are consistent with those of resistivity measurements. Additionally, the SC volume fraction at 2 K is estimated to be about 100% for all samples, indicating bulk superconductivity in H$_x$-FeSe single crystals. As shown in Fig. S3 in the Supplementary Material, $J_c$ reaches a value of $1 \times 10^6$ A/cm$^2$ at zero field ($T = 5$ K), which is much higher than that of the FeSe single crystal, indicating the bulk nature of superconductivity.

To confirm that our findings are intrinsic and sample independent, one piece of H$_x$-FeSe single crystal with a single 44 K SC phase (SP20, sample has protonated for 20 days) was selected to study the evolution of $T_c$ in the deprotonation process, as shown in Figs. 2(e)–(g).

The 44 K SC phase is rather robust against deprotonation, which exhibits a sharp transition until the fifth day of protonation [see Fig. 2(e)]. Extending to longer deprotonation time, the SC transition at 44 K broadens and is accompanied by the reappearance of the 34 K and 25 K SC phases [see Fig. 2(f)]. By day 13, only a SC transition at 25 K is visible after the total suppression of the 44 K and 34 K SC phases. This 25 K SC transition survives up to the 36th day, although it becomes broader. Finally, the nematic and SC transitions in pristine FeSe are restored after the $H^+$ is completely deintercalated at the 40th day [see Fig. 2(g)]. The observation of discrete SC phases in both the protonation and deprotonation processes with an opposite trend in $T_c$ proves the reliability of our results. In addition, it is noted that the normal state transport behavior is also significantly changed with carrier concentration, which will be discussed later.

Summaries showing how the superconductivity of $H_x$-FeSe single crystals evolve with protonation/deprotonation time are presented in Fig. 2(d) and (h). Clearly different from dome-shaped SC phase diagrams established in Fe-based superconductors, discrete SC phases with step-like behavior are observed for both the protonation and deprotonation processes, analogous with (Li/Na)$_x$FeSe flakes via solid ionic gating [35] and $A$-NH$_3$ ($A$ = Na and K) intercalated FeSe systems [7,10]. Note that, besides the two previously reported SC phases with $T_c$ values exceeding 30 K and 40 K in gate-voltage-modulated (Li/Na)$_x$FeSe flakes and other intercalated FeSe systems, an additional 25 K SC phase is detected here for the first time. The origin of this discrepancy is still unclear at present. Nevertheless, our experimental results further prove that the discrete SC phases are intrinsic in intercalated FeSe superconductors. Moreover, the SC transition at 34 K does not exist alone but is always accompanied by a SC transition at 25 K or 44 K, signifying a narrow doping window for the 34 K SC phase. A similar phenomenon has also been observed in K/NH$_3$ co-intercalated FeSe samples, although there is an opposite trend for $T_c$ with electron doping between K/NH$_3$- and $H^+$-intercalated FeSe systems [7].

Previous studies have proposed that the $T_c$ in intercalated FeSe increases with increasing interlayer distance, $d$, until reaching a threshold value of around 9 Å, above which $T_c$ saturates [14,15]. To check this point, a series of $H_x$-FeSe single crystals with different protonation time were placed in a powder X-ray diffractometer and the results are shown in Fig. 3(a). The positions of all (00$l$) reflections exhibit no obvious shift after protonation, indicating that the intercalated $H^+$ has a negligible effect on $d$ (see the inset of Fig. 3(a)) due to its small ionic radius [37]. Moreover, it has been found that the lattice parameter $a$ is robust to the intercalation in intercalated FeSe samples, even though the lattice parameter $c$ can be expanded significantly. We thus believe that the lattice parameter $a$ in $H_x$-FeSe should also remain unchanged. To examine our speculation, two representative $H_x$-FeSe single crystals with protonation time of 8 and 20 days (SP8 and SP20) were selected for single crystal X-ray diffraction measurements. The results of structure refinements are summarized in Tables S1 and S2 in the Supplementary Material. Results show that both the refined lattice parameters $a$ and $c$ are comparable to those of pure FeSe single crystal. Therefore, we can definitely conclude that the intercalation $H^+$ has negligible effect on the protonated FeSe lattice structure. We summarize $T_c$ values of intercalated FeSe samples reported thus far and plot them as a function of $d$ in Fig. 3(b) (data are taken from the previous reports [6,7,9,11,14-17,35,44,45]). Clearly, $T_c$ can be significantly enhanced with increasing carrier concentration in $H_x$-FeSe single crystals, albeit with no change of $d$. Our results definitively indicate that the carrier concentration plays an essential role for achieving high-$T_c$ superconductivity in intercalated FeSe superconductors, while the interlayer distance is not a primary factor. This argument is also supported by previous experimental results. For example, in $K_{0.3}(NH_3)_{0.47}Fe_2Se_2$, if the electron doping level is fixed, $T_c$ remains unchanged even while the $d$ value is considerably decreased from 7.78 to 7.14 Å by deintercalating the $NH_3$ molecule [see closed red circles in Fig. 3(b)] [7]. Recent density functional theory calculations on the $Li_x(NH_2)_y(NH_3)_zFe_2Se_2$ family [46] and experimental studies on Na/DA- and Sr/DA-FeSe systems [represented by closed black squares in Fig. 3(b)] [16] suggest that $T_c$ can be further enhanced with increasing carrier concentrations when $d > 9$ Å.

To gain more insight into the microscopic mechanism of high-$T_c$ superconductivity, as well as into the discrete SC phases, it is of great importance to study normal state properties, which are generally correlated with the SC state in unconventional superconductors. We analyzed the normal state resistivity and attempted to establish a connection with the evolution of $T_c$. The normalized resistivity is fitted with a power law, $\rho(T)/\rho_{200 K} = \rho_0 + AT^\alpha$, where $\rho_0$ is the normalized residual resistivity and $\alpha$ is the resistivity exponent [47-51]. The fits are performed in the temperature region of $T_c$ to $T_s$ or 100 K, and the fitting curves are represented by black dotted lines in Figs. 2(a)–(c) and 2(e)–(g). The obtained $\alpha$ for both the protonation and deprotonation processes are summarized in the phase diagram shown in Fig. 5(a) and (b). As can be seen from Fig. 5(a), the non-Fermi-liquid behavior with $\alpha \approx 1$ occurs inside the nematic phase. After the nematic phase is completely suppressed, the value of $\alpha$ abruptly increases to ~ 1.5 and keeps a nearly constant value within a certain range in the $T_{c2}$ SC phase. Then, $\alpha$ gradually increases until saturating at $\alpha \approx 2$ with further increasing carrier concentration, indicating an evolution from non-Fermi- to Fermi-liquid transport. This Fermi-liquid behavior has been observed in heavily electron-doped FeSe-derived superconductors, such as $(Li_{1-x}Fe_x)OHFeSe$ [52] and $Li_x(NH_3)_yFe_2Se_2$ [45]. An opposite trend of $\alpha$ can also be found in the deprotonation process [see Fig. 5(b)]. The evolution of $\alpha$ can be seen more clearly in a contour plot of the temperature-dependent $\alpha$ extracted from $d\ln(\rho - \rho_0)/d\ln T$ in Fig. 5(a) and (b). The more Fermi-liquid-like behavior at high-$T_c$ phases reflects a reduction in electron correlation with doping, consistent with what is generally found in Fe-based superconductors [53] but different from the anomalous behavior in K-dosed FeSe films, where the correlation strength is enhanced with increased doping [21]. More importantly, the exponent $\alpha$ shows almost identical step-like behavior to that observed in $T_c$, except for a certain range where $T_{c2}$, $T_{c3}$, and $T_{c4}$ SC phases coexist. Therefore, this suggests that the discrete SC phases and the change of exponent $\alpha$ in the normal state resistivity have the same origin; namely, they are intimately related to the carrier concentration.

To proceed, we measured the Hall resistivity to probe band structure information. These results are presented in Fig. 4. As shown in Fig. 4(a), $\rho_{xy}$ of pristine FeSe exhibits obvious nonlinear behavior at low temperatures, i.e., the negative sign of Hall coefficient $R_H$ at low fields, but becoming positive at high fields. This feature is consistent with previous reported results [39,54]. According to the two-band model including an electron band and a hole band, $R_H$ can be expressed as [55]:

$$R_H = \frac{1}{e}\frac{(\mu_h^2 n_h - \mu_e^2 n_e) + (\mu_h\mu_e)^2(\mu_0 H)^2(n_h - n_e)}{(\mu_e n_e + \mu_h n_h)^2 + (\mu_h\mu_e)^2(\mu_0 H)^2(n_h - n_e)^2}.$$

According to the above equation, we can obtain that $R_H = e^{-1}(n_h\mu_h^2 - n_e\mu_e^2)/(n_e\mu_e + n_h\mu_h)^2$ when $\mu_0 H \to 0$, and $R_H = e^{-1} \times 1/(n_h - n_e)$ when $\mu_0 H \to \infty$. Therefore, $R_H$ extracted from the slope of the high-field quasilinear part can be regarded an indication of the effective carrier density (i.e., the Hall number $n_H = 1/eR_H$), as has been adopted in previous studies on FeSe system [33,56,57]. In FeSe at low temperatures, the positive value of $R_H$ (= $e^{-1} \times 1/(n_h - n_e)$) at high fields means that $n_h > n_e$. With protonation, $\rho_{xy}$ of the protonated FeSe shows a good linear dependence on the magnetic field up to 9 T at all temperatures, and $R_H$ remains negative, indicating that the dominant carriers are electron. These results indicate that, with the doping of $H^+$ into the energy band of FeSe, the effective hole concentration is suppressed and the dominant carriers change to electrons. Noticeable, the evolution of the temperature dependence of $R_H$ is abruptly changed in the 25 K SC phase and is accompanied by a sudden sign reversal, from positive to negative at low temperatures[see Fig. 4(e) and (f)], similar to that observed in gate-voltage-modulated FeSe thin flakes [33]. This behavior has been attributed to a dramatic modification in the band structure (i.e., a Lifshitz transition). In addition, it is noted that the $R_H$–$T$ curve of the 25 K SC phase shows nonmonotonic temperature dependence with a dip feature, which mimics that in the (Li,Fe)OHFeSe system (see Fig. 4(e)) [58]. This implies that the electronic structure of the 25 K SC phase is similar to that in (Li,Fe)OHFeSe, in which only electron pockets exist in the Brillouin zone [25,26] and providing further evidence of a Lifshitz transition. Compared with (Li,Fe)OHFeSe [58], the lower value of $T_{c2}$ in $H_x$-FeSe can be attributed to lower carrier concentrations, as revealed by the larger absolute value of $R_H$ shown in Fig. 4(e). As the protonation time further increases, the absolute value of $n_H$ increases monotonously, indicating the effective electron doping by protonation. Simultaneously, two SC phases with $T_{c3} \approx 34$ K and $T_{c4} \approx 44$ K appear. For the 44 K SC phase, the temperature dependence and, even for its absolute value, $R_H$ are quite similar to those in Li-NH$_3$-intercalated FeSe single crystals with $T_c \approx 44.5$ K [45] and gate-voltage-modulated FeSe thin flakes with $T_c \approx 48$ K [33].

According to the Hall effect results and the knowledge on the FS topology in heavily electron doped FeSe-derived superconductors [21-28], we propose a possible physical picture associated with the Lifshitz transition to explain the sudden increase of $T_c$. Some of the electrons fill the hole pockets at the center of the Brillouin zone ($\Gamma$ point) with electron doping, while other electrons are distributed to other electron pockets at the corners ($M$ point). At the critical doping level where a Lifshitz transition occurs, the hole pockets at the $\Gamma$ point vanish and only electron pockets at the $M$ point exist. Based on this picture, we provide schematic representations of the evolution of FS topology during the protonation and deprotonation processes, as shown in Fig. 5(c) and (d), respectively. With electron doping, the Lifshitz transition occurs and leads to the abrupt appearance of the $T_{c2}$ SC phase, as well as the change of $\alpha$. As the electron doping level is further increased, the radii of electron pockets become larger and larger, corresponding to the subsequent $T_{c3}$ and $T_{c4}$ SC phases. For the $T_{c3}$ and $T_{c4}$ SC phases, a common FS topology with only electron pockets in the Brillouin zone corners has been demonstrated in FeSe-derived superconductors such as KFe$_2$Se$_2$ [23,24], (Li$_{0.8}$Fe$_{0.2}$)OHFeSe [25,26], and K-dosed FeSe films [21,22]. The change of carrier concentration is directly related to the density of states, Fermi surface topology, pairing strength, etc., and these parameters determine $T_c$ together. Some mechanisms such as interpocket pairing between electron pockets [59] and orbital-selective pairing [60] may be responsible for the higher $T_{c3}$ and $T_{c4}$ SC phases [35]. Another possible origin is that the intercalated H$^+$ may occupy different sites in the Wyckoff position of FeSe. H$^+$ ions diffuse into FeSe and occupy the position from a low activation energy site. Upon increasing the concentration, H$^+$ ions may occupy different positions of a higher activation energy or form an ordered state, which leading to the discrete SC phases. Further efforts, such as muon spin relaxation measurements, are warranted to detect the position of $H^+$. Besides the position of $H^+$, the evolution of the Fermi topological structure is also intimately related to the discrete SC phases. Angle resolved photoemission spectroscopy measurement is also necessary to directly identify the evolution of the electronic structures, and clarify their relationship with the discrete SC phases in $H^+$ intercalated FeSe single crystals.

# 4  Conclusions

In summary, we have discovered discrete SC phases in $H_x$-FeSe single crystals during both protonation and deprotonation processes. The lattice structure of $H_x$-FeSe stays almost unchanged during the protonation process. These observations demonstrate that the discrete SC phases are intrinsic and intimately related to the carrier concentration in intercalated FeSe superconductors, whereas the FeSe interlayer distance is not a primary factor for achieving high-$T_c$ superconductivity. Moreover, accompanied with the emergence of discrete SC phases, the normal state resistivity evolves from non-Fermi-liquid to Fermi-liquid behavior with the increase of $T_c$. The systematic change in the FS topology induced by electron doping should be responsible for the evolution of both normal state transport properties and superconductivity. The only existence of electron pockets in the Brillouin zone is in favor of the appearance of high-$T_c$ SC phases.

**Acknowledgements**

This work was partly supported by the National Key R&D Program of China (Grant No. 2018YFA0704300), the Strategic Priority Research Program (B) of the Chinese Academy of Sciences (Grant No. XDB25000000), the National Natural Science Foundation of China (Grants No. U1932217 and No. 11674054), the JSPS KAKENHI (Grants No. 19H02177, No. JP20H05164, and No. JP19K14661), and the JST-Mirai Program (Grant No. JPMJMI17A2).

**Conflict of interest**

The authors declare that they have no conflict of interest.

**Reference**

[1] Shibauchi, T.; Hanaguri, T.; Matsuda, Y. Exotic Superconducting States in FeSe-based Materials. *J. Phys. Soc. Jpn.* 2020, *89*, 102002.

[2] Coldea, A. I. Electronic Nematic States Tuned by Isoelectronic Substitution in Bulk $FeSe_{1-x}S_x$. *Front. Phys.* 2021, *8*, 528.

[3] Hsu, F. C.; Luo, J. Y.; Yeh, K. W.; Chen, T. K.; Huang, T. W.; Wu, P. M.; Lee, Y. C.; Huang, Y. L.; Chu, Y. Y.; Yan, D. C.; Wu, M. K. Superconductivity in the PbO-type structure α-FeSe. *Proc. Natl. Acad. Sci. USA* 2008, *105*, 14262.

[4] Sun, J. P.; Matsuura, K.; Ye, G. Z.; Mizukami, Y.; Shimozawa, M.; Matsubayashi, K.; Yamashita, M.; Watashige, T.; Kasahara, S.; Matsuda, Y.; Yan, J. Q.; Sales, B. C.; Uwatoko, Y.; Cheng, J. G.; Shibauchi, T. Dome-shaped magnetic order competing with high-temperature superconductivity at high pressures in FeSe. *Nat. Commun.* 2016, *7*, 12146.

[5] Guo, J.; Jin, S.; Wang, G.; Wang, S.; Zhu, K.; Zhou, T.; He, M.; Chen, X. Superconductivity in the iron selenide $K_xFe_2Se_2$ ($0 \leq x \leq 1.0$). *Phys. Rev. B* 2010, *82*, 180520.

[6] Ying, T. P.; Chen, X. L.; Wang, G.; Jin, S. F.; Zhou, T. T.; Lai, X. F.; Zhang, H.; Wang, W. Y. Observation of superconductivity at 30~46 K in $A_xFe_2Se_2$ (A = Li, Na, Ba, Sr, Ca, Yb, and Eu). *Sci. Rep.* 2012, *2*, 426.

[7] Ying, T.; Chen, X.; Wang, G.; Jin, S.; Lai, X.; Zhou, T.; Zhang, H.; Shen, S.; Wang, W. Superconducting Phases in Potassium-Intercalated Iron Selenides. *J. Am. Chem. Soc.* 2013, *135*, 2951.

[8] Scheidt, E. W.; Hathwar, V. R.; Schmitz, D.; Dunbar, A.; Scherer, W.; Mayr, F.; Tsurkan, V.; Deisenhofer, J.; Loidl, A. Superconductivity at $T_c$ = 44 K in $Li_xFe_2Se_2(NH_3)_y$. *Eur. Phys. J. B* 2012, *85*, 279.

[9] Burrard-Lucas, M.; Free, D. G.; Sedlmaier, S. J.; Wright, J. D.; Cassidy, S. J.; Hara, Y.; Corkett, A. J.; Lancaster, T.; Baker, P. J.; Blundell, S. J.; Clarke, S. J. Enhancement of the superconducting transition temperature of FeSe by intercalation of a molecular spacer layer. *Nat. Mater.* 2013, *12*, 15.

[10] Zheng, L.; Miao, X.; Sakai, Y.; Izumi, M.; Goto, H.; Nishiyama, S.; Uesugi, E.; Kasahara, Y.; Iwasa, Y.; Kubozono, Y. Emergence of Multiple Superconducting Phases in $(NH_3)_yM_xFeSe$ (M: Na and Li). *Sci. Rep.* 2015, *5*, 12774.

[11] Guo, J.; Lei, H.; Hayashi, F.; Hosono, H. Superconductivity and phase instability of $NH_3$-free Na-intercalated $FeSe_{1-z}S_z$. *Nat. Commun.* 2014, *5*, 4756.

[12] Lu, X. F.; Wang, N. Z.; Wu, H.; Wu, Y. P.; Zhao, D.; Zeng, X. Z.; Luo, X. G.; Wu, T.; Bao, W.; Zhang, G. H.; Huang, F. Q.; Huang, Q. Z.; Chen, X. H. Coexistence of superconductivity and antiferromagnetism in $(Li_{0.8}Fe_{0.2})OHFeSe$. *Nat. Mater.* 2015, *14*, 325.

[13] Krzton-Maziopa, A.; Pomjakushina, E. V.; Pomjakushin, V. Y.; von Rohr, F.; Schilling, A.; Conder, K. Synthesis of a new alkali metal-organic solvent intercalated iron selenide superconductor with $T_c$ approximately 45 K. *J. Phys.: Condens. Matter* 2012, *24*, 382202.

[14] Hatakeda, T.; Noji, T.; Sato, K.; Kawamata, T.; Kato, M.; Koike, Y. New Alkali-Metal- and 2-Phenethylamine-Intercalated Superconductors $A_x(C_8H_{11}N)_yFe_{1-z}Se$ (A = Li, Na) with the Largest Interlayer Spacings and $T_c$ ~ 40 K. *J. Phys. Soc. Jpn.* 2016, *85*, 103702.

[15] Hosono, S.; Noji, T.; Hatakeda, T.; Kawamata, T.; Kato, M.; Koike, Y. New Superconducting Phase of $Li_x(C_6H_{16}N_2)_yFe_{2-z}Se_2$ with $T_c$= 41 K Obtained through the Post-Annealing. *J. Phys. Soc. Jpn.* 2016, *85*, 013702.

[16] Hayashi, F.; Lei, H.; Guo, J.; Hosono, H. Modulation effect of interlayer spacing on the superconductivity of electron-doped FeSe-based intercalates. *Inorg. Chem.* 2015, *54*, 3346.

[17] Shi, M. Z.; Wang, N. Z.; Lei, B.; Shang, C.; Meng, F. B.; Ma, L. K.; Zhang, F. X.; Kuang, D. Z.; Chen, X. H. Organic-ion-intercalated FeSe-based superconductors. *Phys. Rev. Mater.* 2018, *2*.

[18] Rendenbach, B.; Hohl, T.; Harm, S.; Hoch, C.; Johrendt, D. Electrochemical Synthesis and Crystal Structure of the Organic Ion Intercalated Superconductor (TMA)$_{0.5}$Fe$_2$Se$_2$ with $T_c$ = 43 K. *J. Am. Chem. Soc.* 2021, *143*, 3043.

[19] Wang, Q.-Y.; Li, Z.; Zhang, W.-H.; Zhang, Z.-C.; Zhang, J.-S.; Li, W.; Ding, H.; Ou, Y.-B.; Deng, P.; Chang, K.; Wen, J.; Song, C.-L.; He, K.; Jia, J.-F.; Ji, S.-H.; Wang, Y.-Y.; Wang, L.-L.; Chen, X.; Ma, X.-C.; Xue, Q.-K. Interface-Induced High-Temperature Superconductivity in Single Unit-Cell FeSe Films on SrTiO$_3$. *Chin. Phys. Lett.* 2012, *29*, 037402.

[20] Ge, J. F.; Liu, Z. L.; Liu, C.; Gao, C. L.; Qian, D.; Xue, Q. K.; Liu, Y.; Jia, J. F. Superconductivity above 100 K in single-layer FeSe films on doped SrTiO$_3$. *Nat. Mater.* 2015, *14*, 285.

[21] Wen, C. H.; Xu, H. C.; Chen, C.; Huang, Z. C.; Lou, X.; Pu, Y. J.; Song, Q.; Xie, B. P.; Abdel-Hafiez, M.; Chareev, D. A.; Vasiliev, A. N.; Peng, R.; Feng, D. L. Anomalous correlation effects and unique phase diagram of electron-doped FeSe revealed by photoemission spectroscopy. *Nat. Commun.* 2016, *7*, 10840.

[22] Miyata, Y.; Nakayama, K.; Sugawara, K.; Sato, T.; Takahashi, T. High-temperature superconductivity in potassium-coated multilayer FeSe thin films. *Nat. Mater.* 2015, *14*, 775.

[23] Shein, I. R.; Ivanovskii, A. L. Electronic structure and Fermi surface of new K intercalated iron selenide superconductor K$_x$Fe$_2$Se$_2$. *Phys. Lett. A* 2011, *375*, 1028.

[24] Qian, T.; Wang, X. P.; Jin, W. C.; Zhang, P.; Richard, P.; Xu, G.; Dai, X.; Fang, Z.; Guo, J. G.; Chen, X. L.; Ding, H. Absence of a holelike fermi surface for the iron-based K$_{0.8}$F$_{1.7}$Se$_2$ superconductor revealed by angle-resolved photoemission spectroscopy. *Phys. Rev. Lett.* 2011, *106*, 187001.

[25] Zhao, L.; Liang, A.; Yuan, D.; Hu, Y.; Liu, D.; Huang, J.; He, S.; Shen, B.; Xu, Y.; Liu, X.; Yu, L.; Liu, G.; Zhou, H.; Huang, Y.; Dong, X.; Zhou, F.; Liu, K.; Lu, Z.; Zhao, Z.; Chen, C.; Xu, Z.; Zhou, X. J. Common electronic origin of superconductivity in (Li,Fe)OHFeSe bulk superconductor and single-layer FeSe/SrTiO$_3$ films. *Nat. Commun.* 2016, *7*, 10608.

[26] Niu, X. H.; Peng, R.; Xu, H. C.; Yan, Y. J.; Jiang, J.; Xu, D. F.; Yu, T. L.; Song, Q.; Huang, Z. C.; Wang, Y. X.; Xie, B. P.; Lu, X. F.; Wang, N. Z.; Chen, X. H.; Sun, Z.; Feng, D. L. Surface electronic structure and isotropic superconducting gap in(Li$_{0.8}$Fe$_{0.2}$)OHFeSe. *Phys. Rev. B* 2015, *92*, 060504(R).

[27] He, S.; He, J.; Zhang, W.; Zhao, L.; Defa Liu; Xu Liu; Mou, D.; Ou, Y.-B.; Wang, Q.-Y.; Li, Z.; Wang, L.; Peng, Y.; Liu, Y.; Chen1, C.; Yu, L.; Liu, G.; Dong, X.; Zhang, J.; Chen, C.; Xu, Z.; Chen, X.; Ma, X.; Xue, Q.; Zhou, X. J. Phase diagram and electronic indication of high-temperature superconductivity at 65K in single-layer FeSe film. *Nat. Mater.* 2013, *12*, 605.

[28] Liu, D.; Zhang, W.; Mou, D.; He, J.; Ou, Y. B.; Wang, Q. Y.; Li, Z.; Wang, L.; Zhao, L.; He, S.; Peng, Y.; Liu, X.; Chen, C.; Yu, L.; Liu, G.; Dong, X.; Zhang, J.; Chen, C.; Xu, Z.; Hu, J.; Chen, X.; Ma, X.; Xue, Q.; Zhou, X. J. Electronic origin of high-temperature superconductivity in single-layer FeSe superconductor. *Nat. Commun.* 2012, *3*, 931.

[29] Ye, J. T.; Inoue, S.; Kobayashi, K.; Kasahara, Y.; Yuan, H. T.; Shimotani, H.; Iwasa, Y. Liquid-gated interface superconductivity on an atomically flat film. *Nat. Mater.* 2010, *9*, 125.

[30] Ueno, K.; Nakamura, S.; Shimotani, H.; Ohtomo, A.; Kimura, N.; Nojima, T.; Aoki, H.; Iwasa, Y.; Kawasaki, M. Electric-field-induced superconductivity in an insulator. *Nat. Mater.* 2008, *7*, 855.

[31] Shiogai, J.; Ito, Y.; Mitsuhashi, T.; Nojima, T.; Tsukazaki, A. Electric-field-induced superconductivity in electrochemically etched ultrathin FeSe films on SrTiO$_3$ and MgO. *Nat. Phys.* 2015, *12*, 42.

[32] Hanzawa, K.; Sato, H.; Hiramatsu, H.; Kamiya, T.; Hosono, H. Electric field-induced superconducting transition of insulating FeSe thin film at 35 K. *Proc. Natl. Acad. Sci. U.S.A.* 2016, *113*, 3986.

[33] Lei, B.; Cui, J. H.; Xiang, Z. J.; Shang, C.; Wang, N. Z.; Ye, G. J.; Luo, X. G.; Wu, T.; Sun, Z.; Chen, X. H. Evolution of High-Temperature Superconductivity from a Low-$T_c$ Phase Tuned by Carrier Concentration in FeSe Thin Flakes. *Phys. Rev. Lett.* 2016, *116*, 077002.

[34] Lei, B.; Wang, N. Z.; Shang, C.; Meng, F. B.; Ma, L. K.; Luo, X. G.; Wu, T.; Sun, Z.; Wang, Y.; Jiang, Z.; Mao, B. H.; Liu, Z.; Yu, Y. J.; Zhang, Y. B.; Chen, X. H. Tuning phase transitions in FeSe thin flakes by field-effect transistor with solid ion conductor as the gate dielectric. *Phys. Rev. B* 2017, *95*, 020503.

[35] Ying, T. P.; Wang, M. X.; Wu, X. X.; Zhao, Z. Y.; Zhang, Z. Z.; Song, B. Q.; Li, Y. C.; Lei, B.; Li, Q.; Yu, Y.; Cheng, E. J.; An, Z. H.; Zhang, Y.; Jia, X. Y.; Yang, W.; Chen, X. H.; Li, S. Y. Discrete Superconducting Phases in FeSe-Derived Superconductors. *Phys. Rev. Lett.* 2018, *121*, 207003.

[36] Jiang, X.; Qin, M.; Wei, X.; Feng, Z.; Ke, J.; Zhu, H.; Chen, F.; Zhang, L.; Xu, L.; Zhang, X.; Zhang, R.; Wei, Z.; Xiong, P.; Liang, Q.; Xi, C.; Wang, Z.; Yuan, J.; Zhu, B.; Jiang, K.; Yang, M.; Wang, J.; Hu, J.; Xiang, T.; Leridon, B.; Yu, R.; Chen, Q.; Jin, K.; Zhao, Z. Enhancement of Superconductivity Linked with Linear-in-Temperature/Field Resistivity in Ion-Gated FeSe Films. arXiv:2103.06512.

[37] Cui, Y.; Zhang, G.; Li, H.; Lin, H.; Zhu, X.; Wen, H.-H.; Wang, G.; Sun, J.; Ma, M.; Li, Y.; Gong, D.; Xie, T.; Gu, Y.; Li, S.; Luo, H.; Yu, P.; Yu, W. Protonation induced high-$T_c$ phases in iron-based superconductors evidenced by NMR and magnetization measurements. *Sci. Bull.* 2018, *63*, 11.

[38] Böhmer, A. E.; Hardy, F.; Eilers, F.; Ernst, D.; Adelmann, P.; Schweiss, P.; Wolf, T.; Meingast, C. Lack of coupling between superconductivity and orthorhombic distortion in stoichiometric single-crystalline FeSe. *Phys. Rev. B* 2013, *87*, 180505(R).

[39] Sun, Y.; Pyon, S.; Tamegai, T. Electron carriers with possible Dirac-cone-like dispersion in FeSe$_{1-x}$S$_x$ (x=0 and 0.14) single crystals triggered by structural transition. *Phys. Rev. B* 2016, *93*, 104502.

[40] Lu, N.; Zhang, P.; Zhang, Q.; Qiao, R.; He, Q.; Li, H. B.; Wang, Y.; Guo, J.; Zhang, D.; Duan, Z.; Li, Z.; Wang, M.; Yang, S.; Yan, M.; Arenholz, E.; Zhou, S.; Yang, W.; Gu, L.; Nan, C. W.; Wu, J.; Tokura, Y.; Yu, P. Electric-field control of tri-state phase transformation with a selective dual-ion switch. *Nature* 2017, *546*, 124.

[41] Dolomanov, O. V.; Bourhis, L. J.; Gildea, R. J.; Howard, J. A. K.; Puschmann, H. OLEX2: a complete structure solution, refinement and analysis program. *J. Appl. Cryst.* 2009, *42*, 339.

[42] Sheldrick, G. M. SHELXT - integrated space-group and crystal-structure determination. *Acta Crystallogr., Sect. A: Found. Adv.* 2015, *71*, 3.

[43] Sheldrick, G. M. Crystal structure refinement with SHELXL. *Acta Crystallogr., Sect. C: Struct. Chem.* 2015, *71*, 3.

[44] Fan, X.; Deng, J.; Chen, H.; Zhao, L.; Sun, R.; Jin, S.; Chen, X. Nematicity and superconductivity in orthorhombic superconductor Na$_{0.35}$(C$_3$N$_2$H$_{10}$)$_{0.426}$Fe$_2$Se$_2$. *Phys. Rev. Mater.* 2018, *2*, 114802.

[45] Sun, S.; Wang, S.; Yu, R.; Lei, H. Extreme anisotropy and anomalous transport properties of heavily electron doped Li$_x$(NH$_3$)$_y$Fe$_2$Se$_2$ single crystals. *Phys. Rev. B* 2017, *96*, 064512.

[46] Guterding, D.; Jeschke, H. O.; Hirschfeld, P. J.; Valentí, R. Unified picture of the doping dependence of superconducting transition temperatures in alkali metal/ammonia intercalated FeSe. *Phys. Rev. B* 2015, *91*, 041112(R).

[47] Analytis, J. G.; Kuo, H.-H.; McDonald, R. D.; MarkWartenbe; Rourke, P. M. C.; Hussey, N. E.; Fisher, I. R. Transport near a quantum critical point in BaFe$_2$(As$_{1-x}$P$_x$)$_2$. *Nat. Phys.* 2014, *10*, 194.

[48] Yi, X.; Xing, X.; Qin, L.; Feng, J.; Li, M.; Zhang, Y.; Meng, Y.; Zhou, N.; Sun, Y.; Shi, Z. Hydrothermal synthesis and complete phase diagram of FeSe$_{1-x}$S$_x$ ($0 \leq x \leq 1$) single crystals. *Phys. Rev. B* 2021, *103*, 144501.

[49] Sun, J. P.; Shahi, P.; Zhou, H. X.; Huang, Y. L.; Chen, K. Y.; Wang, B. S.; Ni, S. L.; Li, N. N.; Zhang, K.; Yang, W. G.; Uwatoko, Y.; Xing, G.; Sun, J.; Singh, D. J.; Jin, K.; Zhou, F.; Zhang, G. M.; Dong, X. L.; Zhao, Z. X.; Cheng, J. G. Reemergence of high-Tc superconductivity in the $(Li_{1-x}Fe_x)OHFe_{1-y}Se$ under high pressure. *Nat. Commun.* 2018, *9*, 380.

[50] Kasahara, S.; Shibauchi, T.; Hashimoto, K.; Ikada, K.; Tonegawa, S.; Okazaki, R.; Shishido, H.; Ikeda, H.; Takeya, H.; Hirata, K.; Terashima, T.; Matsuda, Y. Evolution from non-Fermi-to Fermi-liquid transport via isovalent doping in $BaFe_2(As_{1-x}P_x)_2$ superconductors. *Phys. Rev. B* 2010, *81*, 184519.

[51] Bristow, M.; Reiss, P.; Haghighirad, A. A.; Zajicek, Z.; Singh, S. J.; Wolf, T.; Graf, D.; Knafo, W.; McCollam, A.; Coldea, A. I. Anomalous high-magnetic field electronic state of the nematic superconductors $FeSe_{1-x}S_x$. *Phys. Rev. Research* 2020, *2*, 013309.

[52] Yi, X.; Qin, L.; Xing, X.; Lin, B.; Li, M.; Meng, Y.; Xu, M.; Shi, Z. Synthesis of $(Li_{1-x}Fe_x)OHFeSe$ and FeSe single crystals without using selenourea via a hydrothermal method. *J. Phys. Chem. Solids* 2020, *137*, 109207.

[53] Ye, Z. R.; Zhang, Y.; Chen, F.; Xu, M.; Jiang, J.; Niu, X. H.; Wen, C. H. P.; Xing, L. Y.; Wang, X. C.; Jin, C. Q.; Xie, B. P.; Feng, D. L. Extraordinary Doping Effects on Quasiparticle Scattering and Bandwidth in Iron-Based Superconductors. *Phys. Rev. X* 2014, *4*, 031041.

[54] Watson, M. D.; Yamashita, T.; Kasahara, S.; Knafo, W.; Nardone, M.; Béard, J.; Hardy, F.; McCollam, A.; Narayanan, A.; Blake, S. F.; Wolf, T.; Haghighirad, A. A.; Meingast, C.; Schofield, A. J.; v. Löhneysen, H.; Matsuda, Y.; Coldea, A. I.; Shibauchi, T. Dichotomy between the Hole and Electron Behavior in Multiband Superconductor FeSe Probed by Ultrahigh Magnetic Fields. *Phys. Rev. Lett.* 2015, *115*, 027006.

[55] A., S. R., Semiconductors (Cambridge University Press.

[56] Sun, Y.; Taen, T.; Yamada, T.; Pyon, S.; Nishizaki, T.; Shi, Z.; Tamegai, T. Multiband effects and possible Dirac fermions in $Fe_{1+y}Te_{0.6}Se_{0.4}$. *Phys. Rev. B* 2014, *89*, 144512.

[57] Lei, H.; Graf, D.; Hu, R.; Ryu, H.; Choi, E. S.; Tozer, S. W.; Petrovic, C. Multiband effects on *β*-FeSe single crystals. *Phys. Rev. B* 2012, *85*, 094515.

[58] Dong, X.; Jin, K.; Yuan, D.; Zhou, H.; Yuan, J.; Huang, Y.; Hua, W.; Sun, J.; Zheng, P.; Hu, W.; Mao, Y.; Ma, M.; Zhang, G.; Zhou, F.; Zhao, Z. $(Li_{0.84}Fe_{0.16})OHFe_{0.98}Se$ superconductor: Ion-exchange synthesis of large single-crystal and highly two-dimensional electron properties. *Phys. Rev. B* 2015, *92*, 064515.

[59] Khodas, M.; Chubukov, A. V. Interpocket pairing and gap symmetry in Fe-based superconductors with only electron pockets. *Phys. Rev. Lett.* 2012, *108*, 247003.

[60] Nica, E. M.; Yu, R.; Si, Q. Orbital-selective pairing and superconductivity in iron selenides. *npj Quantum Mater.* 2017, *2*, 24.

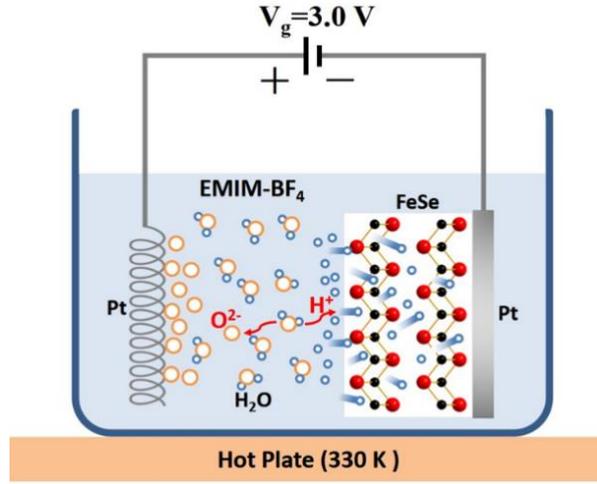

**Fig. 1.** Schematic of the experimental setup used for protonation.

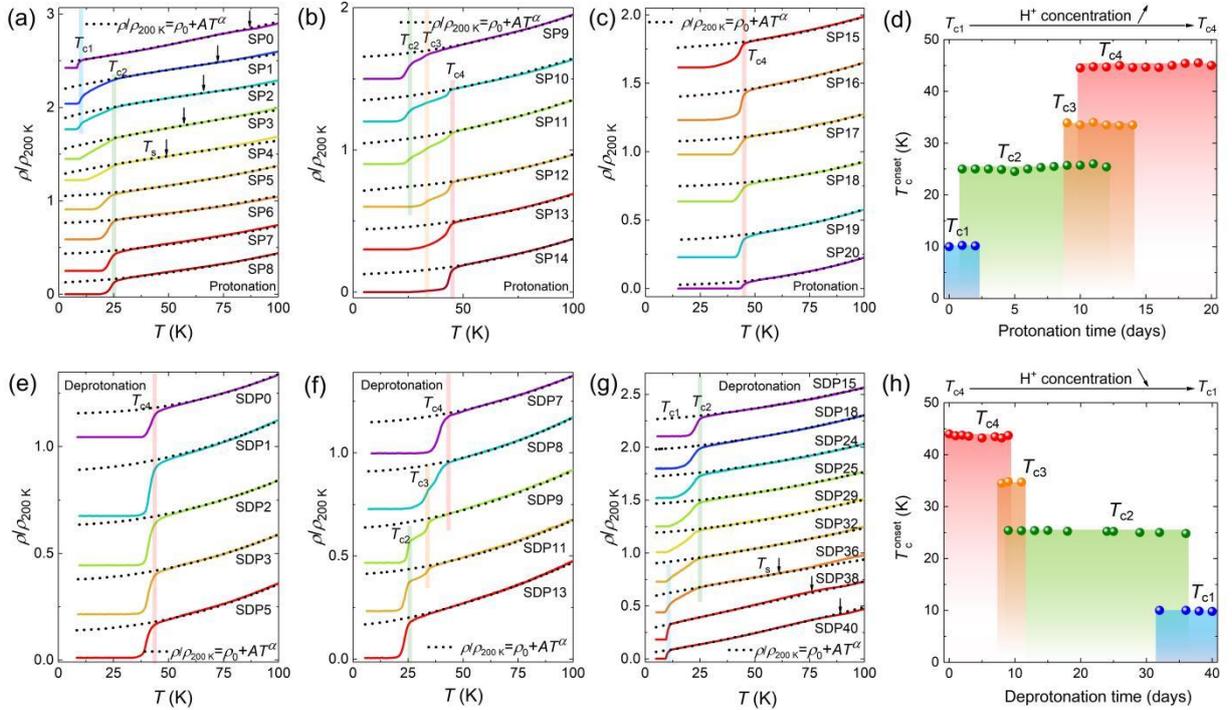

**Figs. 2.** (a–c) Temperature dependence of the normalized resistivity, $\rho(T)/\rho_{200\,K}$, for different $H_x$-FeSe single crystals during the protonation process with different protonation times. (e–g) $\rho(T)/\rho_{200\,K}$ curves for $H_x$-FeSe single crystals that have been protonated for 20 days (SP20, shown in (c)) in the deprotonation process with different deprotonation time. The curves have been vertically shifted for clarity. Black arrows indicate the nematic transition at $T_s$. Thick vertical lines are guides for the eye and highlight onset SC transition temperatures, $T_{cn}$ ($n$ = 1, 2, 3, and 4), for different SC phases. $T_s$ and $T_{cn}$ ($n$ = 1, 2, 3, and 4) are respectively defined as the dip and peak positions in the temperature derivative of $\rho/\rho_{200\,K}$, as shown in Fig. S1. Black

dotted lines represent fits to the formula $\rho(T)/\rho_{200\,K} = \rho_0 + AT^\alpha$ (see text for details). (d) and (h) summarize the evolution of $T_c$ with (d) protonation and (h) deprotonation time.

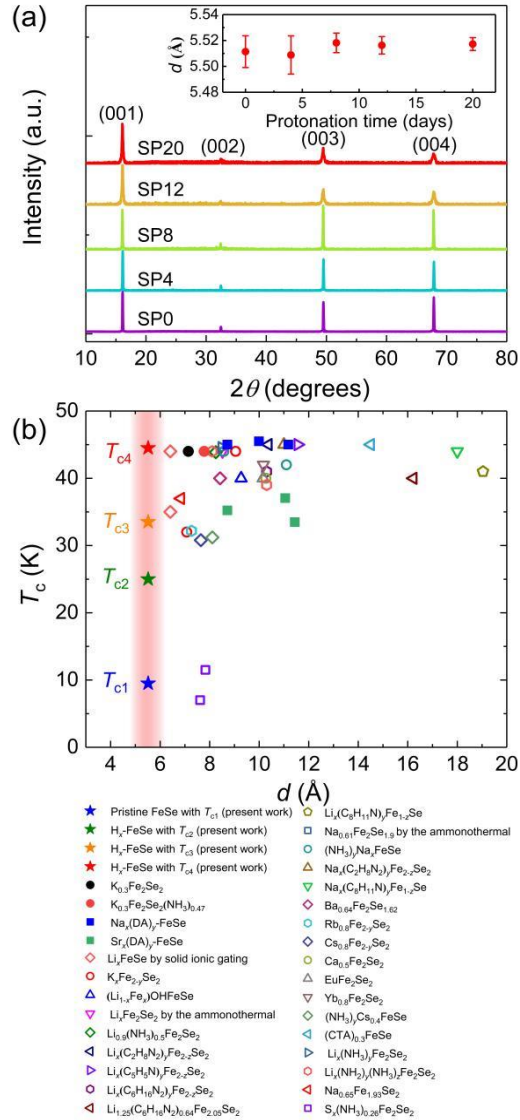

**Fig. 3.** (a) X-ray diffraction patterns of $H_x$-FeSe single crystals with different protonation times performed in a powder X-ray diffractometer. Inset shows the FeSe interlayer distance, $d$, deduced from $2\theta$ values of (00$l$) reflections. (b) Summary of the relationship between $T_c$ and $d$ in a variety of intercalated FeSe samples. The vertical line is a guide for the eye.

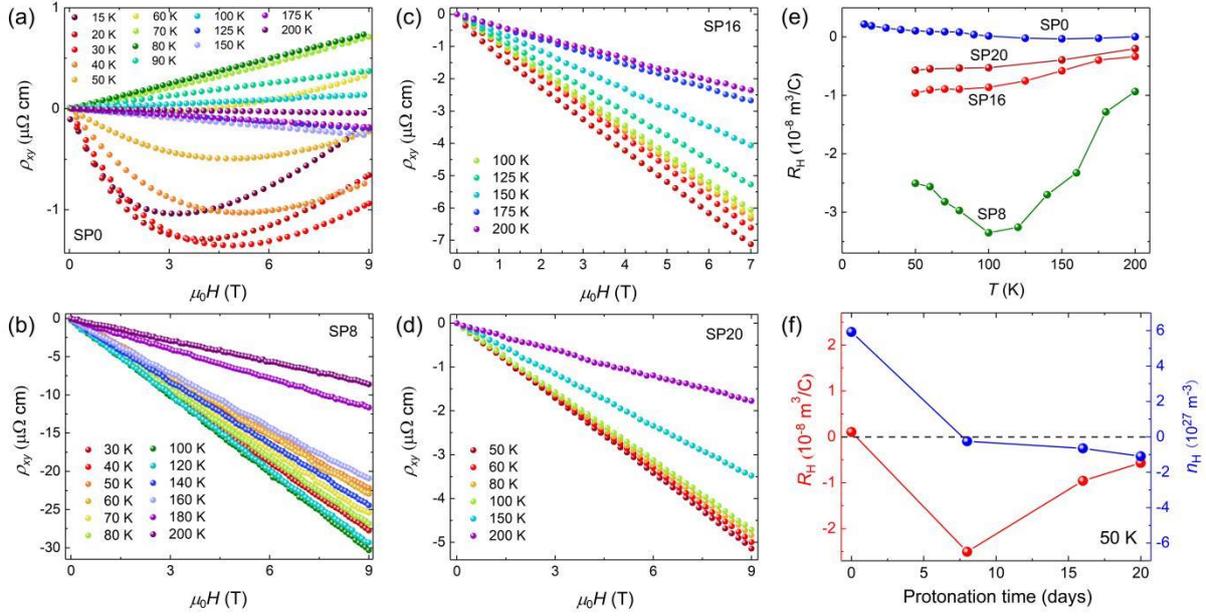

**Figs. 4.** (a–d) Field dependence of Hall resistivity, $\rho_{xy}$, for $H_x$-FeSe single crystals at different temperatures: (a) pristine FeSe single crystal (SP0), $H_x$-FeSe single crystal with a protonation time of (b) eight days (SP8), (c) 16 days (SP16), and (d) 20 days (SP20). (e) Temperature dependence of Hall coefficients of $H_x$-FeSe with different protonation times (for the pristine FeSe, $R_H = \rho_{xy}/H$ was calculated by linear fitting of $\rho_{xy}$ versus $H$ at high magnetic fields). (f) Protonation time dependence of $R_H$ and $n_H$ (Hall number $n_H = 1/eR_H$) at $T = 50$ K.

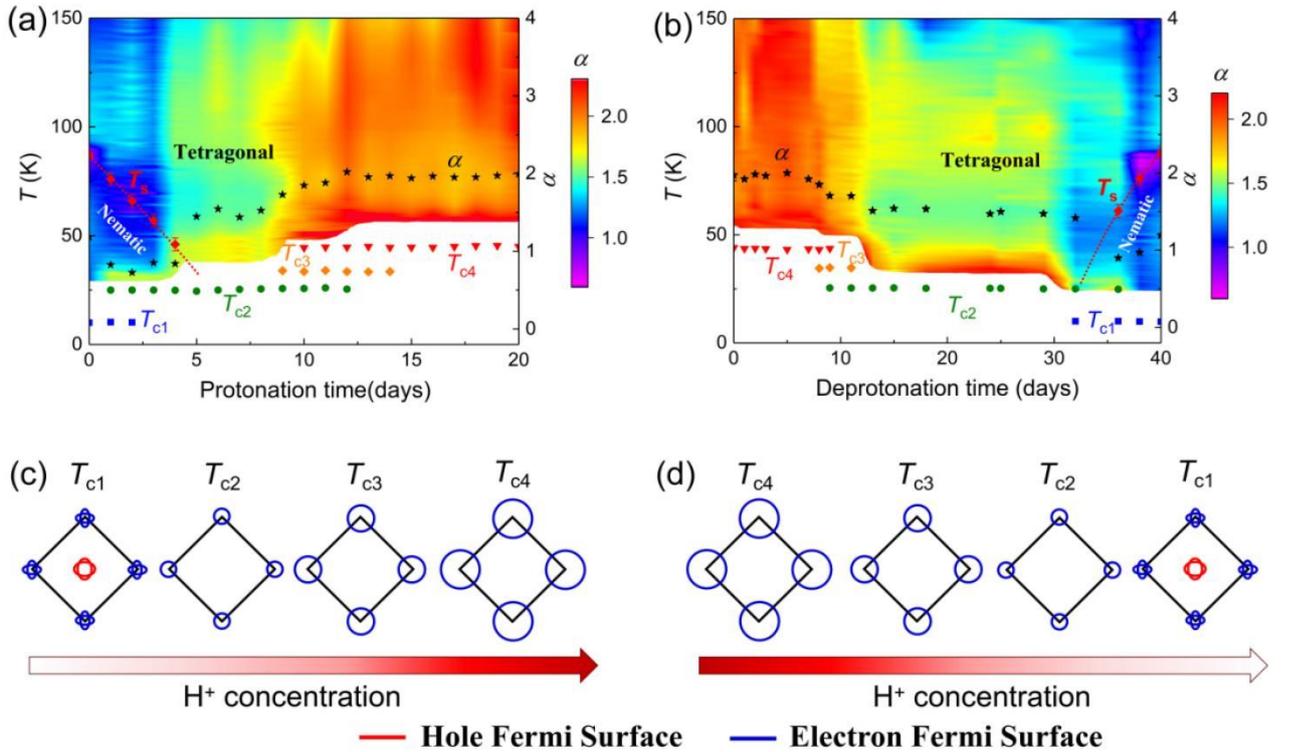

**Figs. 5.** (a and b) Phase diagrams of $H_x$-FeSe single crystals derived from the (a) protonation and (b) deprotonation processes. Color maps represent the temperature dependence of the exponent, $\alpha$, extracted from $d\ln(\rho - \rho_0)/d\ln T$. (c and d) Schematics describing the evolution of FS topology with (c) protonation and (d) deprotonation. Blue and red lines represent the electron and hole FSs, respectively.

# Supplemental Material for "Protonation-induced discrete superconducting phases in bulk FeSe single crystals"

Yan Meng[a,b,1], Xiangzhuo Xing[a, c,1,*], Xiaolei Yi[a,1], Bin Li[d], Nan Zhou[a], Meng Li[a], Yufeng Zhang[a], Wei Wei[a], Jiajia Feng[a], Kensei Terashima[b], Yoshihiko Takano[b], Yue Sun[a,e,*], and Zhixiang Shi[a,*]

[1] *School of Physics, Southeast University, Nanjing 211189, China*

[2] *International Center for Materials Nanoarchitectonics (MANA), National Institute for Materials Science, 1-2-1 Sengen, Tsukuba, Ibaraki 305-0047, Japan*

[3] *School of Physics and Physical Engineering, Qufu Normal University, Shandong 273165, China*

[4] *Information Physics Research Center, Nanjing University of Posts and Telecommunications, Nanjing 210023, China*

[5] *Department of Physics and Mathematics, Aoyama Gakuin University, Sagamihara 252-5258, Japan*

[*]Corresponding authors：xzxing@qfnu.edu.cn, sunyue@phys.aoyama.ac.jp, and zxshi@seu.edu.cn

[1]These authors contributed equally to this work.

## 1. Experimental details of the protonation and deprotonation processes

Pristine FeSe single crystals were grown by a chemical vapor transport (CVT) method described elsewhere [1]. To obtain homogenously protonated samples, the FeSe single crystals used for protonation are cleaved into thin flakes with an average thickness around 15 ±2 μm. Several cleaved crystals are then attached to a negative electrode using silver paint. The distance between the two platinum electrodes is about 15 mm, and an electrostatic voltage is applied. The ionic liquid 1-ethyl-3-methylimidazolium tetrafluoroborate (EMIM-BF$_4$) is used as a medium, and the measurement container is fixed on a thermostatical hot plate. The optimized protonation voltage and temperature are 3 V and 330 K, respectively, which prominently improve the efficiency of H$^+$ intercalation. Previous study has identified that the electrolysis of water (H$_2$O) is the most likely origin for the O$^{2-}$ and H$^+$ ions [2]. By doping ionic liquids with heavy water (D$_2$O), they could trace a clear deuterium signal distributed uniformly inside the H(D)SrCoO$_{2.5}$ [2]. Furthermore, NMR measurement also detected the existence of protons in protonated FeSe [3]. The position of H$^+$ is located in the interstitial sites and randomly absorbed around the anion Se$^{2-}$ [4]. Water molecules are electrolyzed to H$^+$ and O$^{2-}$ ions at optimized protonation voltage and temperature. Benefitting from the high electrochemical window and good conductivity of ionic liquids, H$^+$ ions can exist stably in ionic liquids and maintain a relatively high concentration near the negative electrode. To maintain sufficient residual water in the ionic liquid, a drop of ultrapure water is added to the container per day. This process is essential for obtaining thoroughly protonated FeSe samples. Finally, a series of H$^+$ intercalated FeSe samples with different protonation times, ranging from 1 to 20 days, were obtained.

It is noteworthy that the advantage of this protonation technique is that H$^+$ is nonvolatile and the obtained H$_x$-FeSe samples are sufficiently stable at room temperature, which allows various subsequent measurements [2]. Nevertheless, our experiments found that the protonated samples are still slowly deprotonated (H$^+$ ions can be deintercalated from H$_x$-FeSe) with time at room temperature

(over 2 days) but relatively stable at low temperatures. In this work, an optimal $H_x$-FeSe single crystal with $T_{c4}$ of 44 K that was protonated for 20 days (SP20) is selected to study the evolution of $T_c$ in the deprotonation process, which deprotonates gradually into the subsequent $T_{c3}$, $T_{c2}$, and $T_{c1}$ SC phases. It is worth mentioning that, in order to systematically study the $T_c$ evolution in the deprotonation process, the sample was always kept below 200 K to reduce the rate of deprotonation. The $T_{c4}$ phase was found to be relatively stable for more than 1 week and then deprotonates into lower $T_{c3}$ and $T_{c2}$ phases. $T_{c3}$ is an unstable intermediate phase that quickly disappears after a few days. The $T_{c2}$ phase seems to be much more stable, even if kept at room temperature.

Table S1 Details of single-crystal X-ray diffraction data collection and structure refinement for $H_x$-FeSe with the protonation time of 8 days (SP8).

| Parameter | Value |
| --- | --- |
| Formula | FeSe |
| Formula weight | 134.81 |
| Temperature/K | 150.15 |
| Crystal system | Tetragonal |

| Space group | P4/nmm |
|---|---|
| Lattice parameters | $a=b=3.7654(6)$ Å, $c=5.5078(14)$ Å, $\alpha=\beta=\gamma=90°$ |
| Volume/Å$^3$ | 78.09(3) |
| Z | 2 |
| $\rho_{calc}$ g/cm$^3$ | 5.733 |
| $\mu$/mm$^{-1}$ | 32.270 |
| F(000) | 120.0 |
| Crystal size/mm$^3$ | 0.35 × 0.18 × 0.03 |
| Radiation | Mo K$\alpha$ ($\lambda = 0.71073$) |
| 2θ range for data collection/ ° | 7.398 to 52.812 |
| Index ranges | $-4 \leq h \leq 4$, $-4 \leq k \leq 3$, $-6 \leq l \leq 6$ |
| Reflections collected | 635 |
| Independent reflections | 67 [$R_{int} = 0.0707$, $R_{sigma} = 0.0463$] |
| Data/restraints/parameters | 67/0/6 |
| Goodness-of-fit on F$^2$ | 1.334 |
| Final R indexes [I>=2σ (I)] | $R_1 = 0.0405$, $wR_2 = 0.1326$ |
| Final R indexes [all data] | $R_1 = 0.0420$, $wR_2 = 0.1345$ |
| Largest diff. peak/hole / e Å$^{-3}$ | 0.98/-0.95 |

Table S2 Details of single-crystal X-ray diffraction data collection and structure refinement for H$_x$-FeSe with the protonation time of 20 days (SP20).

| Parameter | Value |
|---|---|
| Formula | FeSe |
| Formula weight | 134.81 |
| Temperature/K | 150.15 |
| Crystal system | tetragonal |
| Space group | P4/nmm |
| Lattice parameters | a=b=3.7716(11) Å, c=5.511(3) Å, α=β=γ=90 ° |
| Volume/Å$^3$ | 78.39(6) |
| Z | 2 |
| $\rho_{calc}$g/cm$^3$ | 5.711 |
| μ/mm$^{-1}$ | 32.145 |
| F(000) | 120.0 |
| Crystal size/mm$^3$ | 0.22×0.18×0.02 |
| Radiation | Mo Kα (λ = 0.71073) |
| 2θ range for data collection/ ° | 14.82 to 52.73 |
| Index ranges | −4 ≤ h ≤ 4, −4 ≤ k ≤ 4, −6 ≤ l ≤ 6 |
| Reflections collected | 462 |
| Independent reflections | 64 [$R_{int}$ = 0.0705, $R_{sigma}$ = 0.0975] |
| Data/restraints/parameters | 64/0/6 |
| Goodness-of-fit on F$^2$ | 1.328 |
| Final R indexes [I>=2σ (I)] | $R_1$ = 0.0421, w$R_2$ = 0.1209 |
| Final R indexes [all data] | $R_1$ = 0.0441, w$R_2$ = 0.1227 |
| Largest diff. peak/hole / e Å$^{-3}$ | 1.74/-1.22 |

## 2. Determination of the transition temperatures of $T_c$ and $T_s$

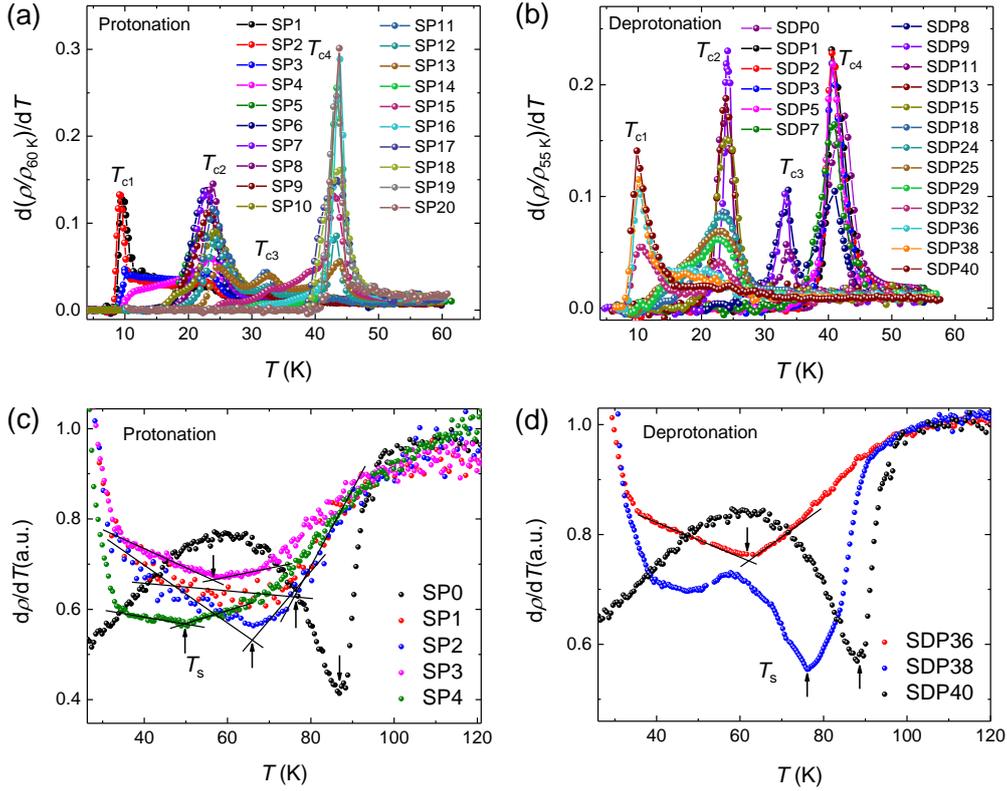

Fig. S1 Derivatives of the normalized temperature dependence of resistivity, d$\rho$/d$T$, near superconducting transitions in the (a) protonation and (b) deprotonation processes. d$\rho$/d$T$ near the nematic transition in the (c) protonation and (d) deprotonation processes. Arrows denote the nematic transition at $T_s$.

## 3. Magnetization characterization

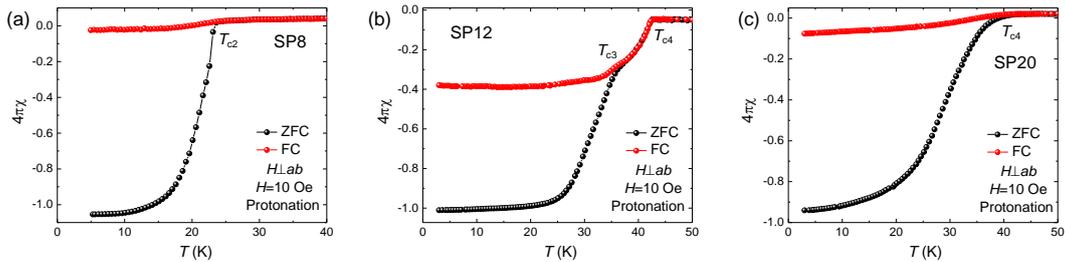

Fig. S2 Temperature dependence of the magnetic susceptibility of the discrete superconducting phases. The magnetization was measured under zero-field cooling (ZFC) and field cooling (FC) under an applied magnetic field of 10 Oe with $H \perp ab$ axis. (a) Pure $T_{c2}$ SC phase with a protonation time of 8 days (SP8), the size of SP8 is about 810 μm × 540 μm × 15 μm and corresponding demagnetizing factor N is calculated as.0.934. (b) $T_{c3}$ SC phase accompanied by the $T_{c4}$ SC phase with a protonation time of 12 days (SP12), the size of SP12 is about 780 μm × 450 μm × 14 μm and corresponding demagnetizing factor N is calculated as.0.931. (c) Optimal $T_{c4}$ SC phase with a protonation time of 20 days (SP20), the size of SP20 is about 720 μm × 410 μm × 13 μm and corresponding demagnetizing factor N is calculated as 0.930.

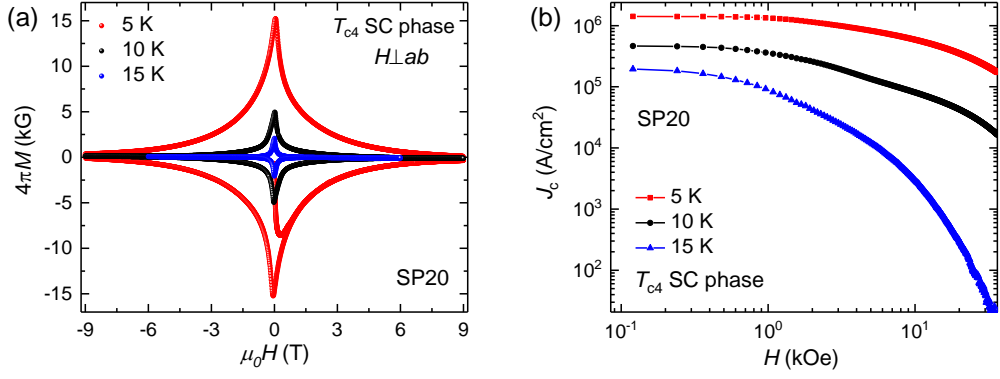

Fig. S3 (a) Magnetization hysteresis loops (MHLs) of the optimal $T_{c4}$ SC phase (SP20) at $T$ = 5 K, 10 K, and 15 K for $H \perp ab$. (b) Corresponding magnetic field dependence of critical current densities, $J_c$, derived from the Bean model.

The optimal $T_{c4}$ SC phase of $H_x$-FeSe with a protonation time of 20 days was selected for magnetization measurements. We calculated the field dependence of $J_c$ from the MHLs based on the Bean model: $J_c = 20\Delta M/[a(1 − a/3b)]$, where $\Delta M$ is $M_{down} − M_{up}$, $M_{up}$ [emu/cm$^3$] and $M_{down}$ [emu/cm$^3$] are the magnetization when sweeping fields up and down, respectively, and $a$ [cm] and $b$ [cm] are sample widths ($a < b$) [5]. The derived $J_c$ as a function of magnetic field is shown in Fig. S3 (b). $J_c$ reaches a value of $1 \times 10^6$ A/cm$^2$ at zero field ($T$ = 5 K), which is much higher than that of the FeSe single crystal [6], indicating the bulk nature of superconductivity.

## References

[1]. Kasahara, S.; Watashige, T.; Hanaguri, T.; Kohsaka, Y.; Yamashita, T.; Shimoyama, Y.; Mizukami, Y.; Endo, R.; Ikeda, H.; Aoyama, K.; Terashima, T.; Uji, S.; Wolf, T.; von Lohneysen, H.; Shibauchi, T.; Matsuda, Y., Field-induced superconducting phase of FeSe in the BCS-BEC cross-over. *Proc. Natl. Acad. Sci. USA* **2014**, 111 (46), 16309-13.

[2]. Lu, N.; Zhang, P.; Zhang, Q.; Qiao, R.; He, Q.; Li, H. B.; Wang, Y.; Guo, J.; Zhang, D.; Duan, Z.; Li, Z.; Wang, M.; Yang, S.; Yan, M.; Arenholz, E.; Zhou, S.; Yang, W.; Gu, L.; Nan, C. W.; Wu, J.; Tokura, Y.; Yu, P., Electric-field control of tri-state phase transformation with a selective dual-ion switch. *Nature* **2017**, 546 (7656), 124-128.

[3]. Cui, Y.; Zhang, G.; Li, H.; Lin, H.; Zhu, X.; Wen, H.-H.; Wang, G.; Sun, J.; Ma, M.; Li, Y.; Gong, D.; Xie, T.; Gu, Y.; Li, S.; Luo, H.; Yu, P.; Yu, W. Protonation induced high-$T_c$ phases in iron-based superconductors evidenced by NMR and magnetization measurements. *Sci. Bull.* **2018**, *63*, 11.

[4]. Cui, Y.; Hu, Z.; Zhang, J.-S.; Ma, W.-L.; Ma, M.-W.; Ma, Z.; Wang, C.; Yan, J.-Q.; Sun, J.-P.; Cheng, J.-G.; Jia, S.; Li, Y.; Wen, J.-S.; Lei, H.-C.; Yu, P.; Ji, W.; Yu, W.-Q., Ionic-Liquid-Gating Induced Protonation and Superconductivity in FeSe, FeSe0.93S0.07, ZrNCl, 1T-TaS2 and Bi2Se3 *. *Chin. Phys. Lett.* **2019**, 36 (7).

[5]. Bean, C. P., Magnetization of High-Field Superconductors. *Rev. Mod. Phys.* **1964**, 36 (1), 31-39.

[6]. Sun, Y.; Pyon, S.; Tamegai, T.; Kobayashi, R.; Watashige, T.; Kasahara, S.; Matsuda, Y.; Shibauchi, T., Critical current density, vortex dynamics, and phase diagram of single-crystal FeSe. *Phys. Rev. B* **2015**, 92 (14).